\documentclass[prd,superscriptaddress,nofootinbib,showpacs,preprint]{revtex4}
\usepackage{graphicx}
\usepackage{dcolumn}
\usepackage{bm}
\usepackage{amssymb}
\usepackage{mathrsfs}
\usepackage{amsmath}
\usepackage{epsfig}
\usepackage[dvips]{color}
\usepackage{hhline}
\usepackage[colorlinks]{hyperref}
\usepackage[normalem]{ulem}

\def\lapp{\mathrel{\rlap{\raise.5ex\hbox{$<$}}
                    {\lower.5ex\hbox{$\sim$}}}}
\def\gapp{\mathrel{\rlap{\raise.5ex\hbox{$>$}}
                    {\lower.5ex\hbox{$\sim$}}}}

\begin{document}
\title{Electromagnetic leptogenesis at the TeV scale}
\author{Debajyoti Choudhury}
\email{debajyoti.choudhury@gmail.com}
\affiliation{Department of Physics and Astrophysics, University of Delhi, Delhi 110 007, India.}
\author{Namit Mahajan}
\email{nmahajan@prl.res.in}
\affiliation{Physical Research Laboratory, Ahmedabad-380009, India}
\author{Sudhanwa Patra}
\email{sudhakar@prl.res.in}
\affiliation{Physical Research Laboratory, Ahmedabad-380009, India}
\author{Utpal Sarkar}
\email{utpal@prl.res.in}
\affiliation{Physical Research Laboratory, Ahmedabad-380009, India}
\affiliation{McDonnell Center for the Space Sciences,
Washington University, St, Louis, MO 63130, USA}

\begin{abstract}

We construct an explicit model implementing electromagnetic
leptogenesis.  In a simple extension of the Standard Model, a discrete
symmetry forbids the usual decays of the right-handed neutrinos, while
allowing for an effective coupling between the left-handed and
right-handed neutrinos through the electromagnetic dipole moment.
This generates correct leptogenesis with resonant enhancement and also
the required neutrino mass via a TeV scale seesaw mechanism.  The
model is consistent with low energy phenomenology and would have
distinct signals in the next generation colliders, and, perhaps even
the LHC.


\end{abstract}

\pacs{\textcolor{blue}{14.60.Pq, 12.15.Hh} \\
\textcolor{black}{Keywords}:~~\textcolor{blue}{Neutrino Mass, Leptogenesis}}
\maketitle
\section{Introduction}\label{sec:intro}

Several recent experiments have cited convincing evidence in favor of
non-zero neutrino masses and mixing. While both could be admitted in
the Standard Model (SM) by the simple expedient of adding right-handed
neutrino fields (omitted, at the inception of the SM, only on account
of the then apparent masslessness of the neutrinos), many theoretical
challenges persist. Indeed, some authors have claimed neutrino masses
to be the evidence of physics beyond the SM. The very smallness of the
masses accompanied by the largeness of one of the mixing angles, as
also several ``anomalies'' that appear periodically are indicative of
the same. Furthermore, it is not inconceivable that these properties
are related to other unexplained puzzles such as the presence of dark
matter and/or dark energy, as also the matter-antimatter
asymmetry in the universe. It is the last aspect that we shall
concentrate on.

The seesaw mechanism\cite{mink} and the associated mechanism of leptogenesis
\cite{yana} are very attractive means to explain the origin of the
small neutrino masses and the baryon asymmetry of the
universe. Leptogenesis provides an elegant mechanism to consistently
address the observed Baryon Asymmetry in the Universe (BAU)
\cite{dukley} in minimal extensions of the SM
\cite{buch}. In standard leptogenesis, at least two of the right
handed neutrinos should be heavy with masses close to the GUT scale
($\sim 10^{15}$ GeV) and their out-of-equilibrium lepton number
violating decay would create a net lepton asymmetry which,
subsequently, would get converted into the observed baryon asymmetry
via the ($B+L$)-violating sphaleron interactions \cite{kuz,bary}. At
the same time, the inclusion of the right handed (Majorana) fields
with lepton number violating Majorana masses can explain the observed
smallness of light neutrinos through the seesaw mechanism.

Although the aforementioned scheme is theoretically very attractive,
it suffers from the lack of direct detectability, e.g. at high-energy
colliders such as the LHC or ILC, or in any other foreseeable
experiment.  This has, naturally, led to efforts towards alternative
routes to leptogenesis. A phenomenologically interesting solution to
this problem may be obtained within the framework of resonant
leptogenesis \cite{res97,filf,res02,res03,res04,res05}.
Characterized by the presence of two (or more)
nearly degenerate (moderately) heavy Majorana neutrinos, in
such scenarios the corrections to the self-energies play a pivotal
role in determining the lepton asymmetry \cite{kuz,liu}. Indeed, if the
mass difference be comparable to their decay widths, the resonant
enhancement could render asymmetries to be as large as ${\cal O}(1)$
\cite{res97,res05}.

Recently, a very interesting possibility of electromagnetic
leptogenesis \cite{bell} has been proposed, wherein the source of CP
violation has been identified with the electromagnetic dipole
moment(s) of the neutrino(s).  As is well known, the electric
neutrality of the neutrino does not preclude its having non-zero
dipole moments.  And while, naively, the presence of a magnetic dipole
moment would seem to call for the presence of a nonzero mass, even
this is not strictly necessary\cite{vol}. Originally mooted to account
for the then apparent correlation of the solar neutrino flux with the
sunspot activity, various schemes have been proposed to generate large
magnetic moments for neutrinos \cite{kim,valle}. It should be noted at
this stage that while Dirac neutrinos can have both direct and
transition magnetic moments, only the latter are allowed for Majorana
neutrinos.  For a collection of neutrino fields of the same chirality,
the most general form of such couplings is given by
$\overline{\nu^c_j} (\mu_{jk} +i \gamma_5 {\cal D}_{jk})
\sigma_{\alpha \beta} \nu_k B^{\alpha \beta}$, where $B^{\alpha
  \beta}$ denotes the $U(1)$ field strength tensor. The magnetic and
electric transition moment matrices, $\mu_{jk}$ and ${\cal D}_{jk}$,
each need to be antisymmetric. For two Majorana neutrinos combining to
give a Dirac particle, the resultant
matrices, clearly, do not suffer from such restrictions.

The aforementioned dimension-five operators are, presumably, generated
by some new physics operative beyond the electroweak scale. With
$CP$-violation being encoded in the structure of the dipole moments,
the decays of heavier neutrinos to lighter ones and a photon, can, in
principle, lead to a lepton asymmetry in the universe.  Although the
proposal is a very interesting one, thus far it has not been
incorporated in any realistic model.  Indeed, the plethora of
constraints suggests that some amount of fine tuning would be
unavoidable in any realistic model.  In this paper, we discuss the
generic problems of any models for electromagnetic leptogenesis and
suggest possible means to evade them. Considering all these issues, we
point out that on allowing some fine tuning and imposing the resonant
condition it may be possible to construct models of resonant
electromagnetic leptogenesis, but that direct detection would need at
least few more years.

\section{The Model}\label{sec:model}
Retaining the gauge symmetry of the SM, we augment the fermion content
by including three right-handed singlet fields $N_{iR}$ and, in
addition, a singly charged vector-like fermion $E$. Also added are a
singly charged scalar ($H^+$) and a pair of Higgs doublets
($\Sigma,~D$). In keeping with our stated paradigm of only one new
scale, all the new masses are assumed to be around a few TeVs. While
it could be arranged that all these masses arise from the vacuum
expectation value of a single scalar field, for simplicity, we
incorporate explicit mass terms.  The entire particle content, along
with the quantum number assignments, is displayed in Table
\ref{tab:SM}.

\begin{table}[!h]
\begin{center}
\caption{Particle content of the proposed Model }
\label{tab:SM}
\begin{tabular}{|c|c|c|c|}
\hline
 &Field & $ SU(3)_C\times SU(2)_L\times U(1)_Y $ & $~Z_2~$ \\
\hline
\hline
Fermions&$Q_L \equiv(u, d)^T_L$        & $(3, 2, 1/6)$         & +  \\
       &$u_R$                          & $(3, 1, 2/3)$       & +   \\
       &$d_R$                          & $(3, 1, -1/3)$      & +    \\
       &$\ell_L \equiv(\nu,~e)^T_L$    & $(1, 2, -1/2)$        & +     \\
       &$e_R$                          & $(1, 1, -1)$          & +      \\
       &$E_L$                          & $(1, 1, -1)$          & $-$      \\
       &$E_R$                          & $(1, 1, -1)$          & $-$      \\
       &$N_R$                          & $(1, 1, 0)$           & $-$       \\
\hline
Scalars&$\Phi$                         & $(1, 2, +1/2)$   & + \\
       &$\Sigma$                       & $(1, 2, +1/2)$   & $-$  \\
       &$D$                            & $(1, 2, +1/2)$   & $-$   \\
       &$H^+$                          & $(1, 1, +1)$     & +    \\
\hline
\hline
\end{tabular}
\end{center}
\end{table}

At this stage, we are faced with a problem generic to
electromagnetic leptogenesis. While the effective 
$\bar N \, \nu \, \gamma$ coupling has to be
present (so as to allow the mandatory $N \to \nu + \gamma$), the
coupling of 
this fermion pair to the SM
Higgs, viz.  $\bar N \ell \Phi$, needs to be highly suppressed on
two counts: $(i)$ to ensure that the light neutrino mass, accruing
from the seesaw mechanism, is not too large, and $(ii)$ to prevent
the $N$ from decaying dominantly to $\ell + \Phi$. While this
could, nominally, be ensured by invoking some symmetry wherein the
photon and the $\Phi$ transform differently, such an assignment
would adversely impact the phenomenology of the charged particles.
We, rather, choose to introduce a discrete $Z_2$ symmetry. All of
the SM particles as well as the charged singlet scalar $H^+$ are
even under this $Z_2$ symmetry, while 
the rest are odd (see Table \ref{tab:SM}).

The $Z_2$ symmetry allows both the (effective)
Majorana mass terms $\bar \nu^c \,
\nu$ and $ \bar N^c \, N$ but the former is precluded if we limit
ourselves to a renormalizable Lagrangian. On the other hand, the
coupling of the neutrinos with the SM Higgs $\Phi$, namely a term of
the form $\bar N \ell \Phi$ is disallowed, thereby preventing an
effective Dirac mass term of the form $\bar N \, \nu$.  More
importantly, it also forbids the magnetic moment term $\bar N \nu
\gamma$.  Each of these can be generated only when the $Z_2$ is
broken. Rather than break it spontaneously, and thereby risk domain
walls, we choose to break it explicitly, but only through a soft
term. While preserving the essential features of the model, this,
then, allows the generation of both Dirac neutrino mass terms as well
as magnetic moments and, thereby, driving resonant leptogenesis
successfully.

While the Yukawa Lagrangian for the quarks remains unchanged from the
SM, that for the leptonic sector can be written as
\begin{eqnarray}
{\cal L}_{\rm Yuk} &\ni& \Bigg[y_H~ \overline{N_R} ~{E_L} H^{+}  +y_\Sigma
\overline{\ell_L} \Sigma E_R + y_D  \overline{\ell_L} D E_R
       \nonumber  \\
&+&h_\Sigma \overline{\ell_L}  \tilde{\Sigma} N_R+
h_D \overline{\ell_L} \tilde{D} N_R+y_e \overline{\ell_L} \Phi e_R +h.c. \bigg]   \nonumber  \\
&+&\Bigg[\frac{1}{2} \overline{(N_R)^C} M_N N_R-M_E \overline{E_R} E_L+h.c.
\Bigg]
\end{eqnarray}
where the last two terms ($M_N, M_E$) represent gauge- and
$Z_2$--invariant bare mass matrices. In the above, $\tilde{\Phi}=i
\sigma_2 \Phi^*$ (similarly for $\tilde{D}$ and $\tilde{\Sigma}$) with
$y_H$, $y_\Sigma$, $y_D$, $h_\Sigma$ and $h_D$ being the Yukawa coupling
matrices.

The scalar potential can be parametrized as
\begin{eqnarray}
V(\Phi, \Sigma, D, H^+) &=& -\mu_\Phi^2|\Phi|^2
+ m_2^2|\Sigma|^2 +m_3^2|D|^2 + m_h^2\,|H|^2 +
\lambda_1|\Phi|^4 + \lambda_2|\Sigma|^4
                   \nonumber \\
&+& \lambda_3|D|^4        +
\lambda_h|H|^4+\lambda_{\Phi H} (\Phi^{\dag}\Phi)\, |H|^2+
\lambda_{D H} (D^{\dag}D)\, |H|^2    \nonumber \\
&+&\lambda_{\Sigma H} (\Sigma^{\dag}\Sigma) |H|^2+ \lambda_{D \Sigma H}
(D^{\dag}\Sigma) |H|^2 +\frac{\lambda_{\Phi \Sigma}}{2}
\left[ (\Phi^{\dag}\Sigma)^2  + h.c. \right]                                                   \nonumber \\
&+& \lambda_{D \Phi} (D^{\dag}\Sigma) (\Phi^{\dag}\Phi)  +
f_1\, (\Phi^{\dag}\Phi)\,(D^{\dag}D) +f_2\, (\Phi^{\dag}\Phi)\,(\Sigma^{\dag}\Sigma)
         \nonumber \\
&+& f_3 \,|\Phi^{\dag}D|^2+ f_4|\Phi^{\dag}\, \Sigma|^2+
f_5\,(D^{\dag}D) \,(\Sigma^{\dag}\Sigma) + f_6|D^{\dag}\, \Sigma|^2                                       \nonumber \\
&+&\left[\mu_s\, \Sigma \cdot D\,(H^+)^\ast + h.c.\right].
\end{eqnarray}
Note that the two fields $D$ and $\Sigma$ are being ascribed a
positive mass-squared each so that $Z_2$ is left unbroken at this
stage. Furthermore, we assume that $m_{2,3}$ are large enough ($\sim
{\cal O}(10 \, {\rm TeV})$) so that the decays $N \to \nu + D/\Sigma$
are kinematically disallowed.

As argued earlier, the $Z_2$ symmetry needs to be broken, and we
achieve this through an explicit soft term.  This has the advantage of
obviating any domain wall problem 
without introducing any qualitative
changes to the rest of the phenomenology. To this end, we 
posit terms of the form
\begin{equation}
V_{soft} = \mu_{soft}^2 \Phi^\dagger D + \cdots
\label{eqn:softbreak}
\end{equation}
without going into 
their origin. It should be noted that although this solves the
problem, in a realistic model one must explain the origin of such
terms, which is somewhat
nontrivial and may plague the model.
The ellipses above denote
other possible terms, such as $\Phi^\dagger \Sigma$ etc. that do not
concern us directly. The scale of the soft symmetry breaking
$\mu_{soft}$ needs to be significantly lower than the electroweak
symmetry breaking scale. This naturally leads to a large gradation in
the vacuum expectation values, namely $\langle D \rangle,\langle
\Sigma \rangle \ll \langle \Phi \rangle$.
The breaking of the $Z_2$ symmetry allows for a non-zero
value of the effective magnetic moment term $\bar N \ell \gamma$,
which is necessary for leptogenesis to go through. Also introduced
is a Dirac mass term $\bar N \, \nu \, \langle D\rangle$.
On the other hand, this breaking now permits the decay
$N \to \nu + \Phi_0$ which proceeds through  the mixing of $\Phi$ with
$D$ and/or $\Sigma$. The twin facts of $N$ being heavy
and $\Phi_0$ being light (unitarity of the SM as well as
consonance with LEP data) implies that this cannot be wished
away on kinematic grounds. Note, however, that this interaction
is suppressed by a factor of $\langle D \rangle~ / ~\langle \Phi \rangle$
and, as we show in the next section, a value commensurate with light
neutrino masses provides adequate suppression.

\section{Neutrino mass}
An exact $Z_2$ symmetry in the Lagrangian prevents the Yukawa term
$\overline{\ell} \Phi N$. On the other hand, the fact of $m_{2,3}^2 > 0$
prevents a vacuum expectation value for both the fields that do couple
to the $\overline{\ell} N$ current, namely $D$ and
$\Sigma$. Consequently, there is no Dirac neutrino mass at this level.
However, once the soft-breaking term of eqn.(\ref{eqn:softbreak}) is
included, the field $D$ may receive a non-zero {\em vev}, despite
positive $m_{2,3}^2$. This, in turn, gives a Dirac mass to the
neutrinos viz.
\begin{equation}
M_{\rm Dirac} = h_D \langle D \rangle = h_D v_D .
\end{equation}
This, together with the Majorana mass term $M_N$ for the heavy
right-chirality fields, gives rise to a light neutrino Majorana mass
via type-I seesaw mechanism, viz.
\begin{equation}
m_\nu=M_{\rm Dirac} M^{-1}_N M_{\rm Dirac} \ .
\end{equation}
For the choice of parameters we are interested in, $M_D \sim 10^{-3}
h_D v ~ \sim 10^{-4} $ GeV ($v = \langle \Phi \rangle \sim 100$~GeV
and $h_D \sim 0.001$). The right-handed neutrinos are heavier than the
SM Higgs scalar. For $M_N \sim $ few TeV, this gives the correct
magnitude of the light neutrino masses, namely $m_\nu \sim
0.1$~eV. The hierarchy of masses could be obtained because of the
different values of the elements of the matrices $M_N$ and $h_D$.
\section{Dipole coupling  between light and heavy neutrinos}
\label{sec:Emdm}
The effective Lagrangian describing the interaction between photon and
the light--heavy ($\bar \nu N$) neutrino-current can, in general, be
parametrized as
\begin{equation}
 \mathcal{L}_\text{EM} =
  \overline{\nu}_{Lj}\, \lambda_{jk}\, \sigma_{\alpha\beta}\,P_R\,
    N_{k} \,F^{\alpha\beta}
         +\text{h.c.}
     \label{eqn_emdm_lag_eff}
\end{equation}
The effective coupling matrix $\lambda_{jk}$ is, in general, a complex
one, and needs to be calculated in terms of the parameters of the
model.
 \begin{figure}[htb]
  \centering
  \includegraphics[width=1.00 \textwidth]{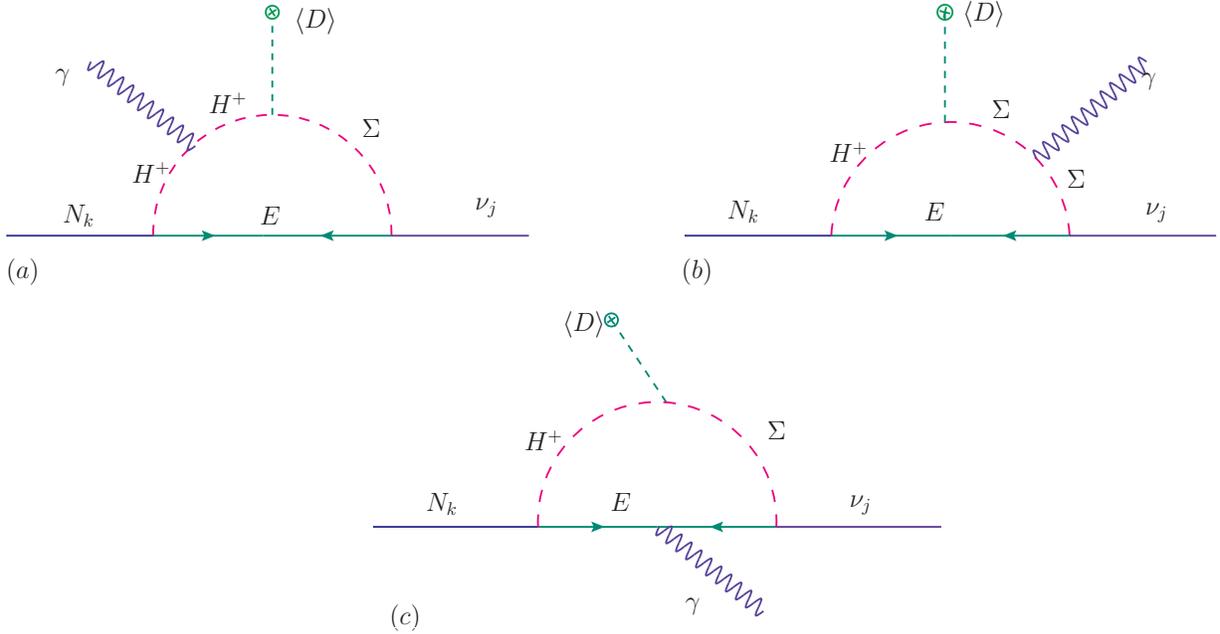}
 \caption{Feynman diagrams leading to the effective EMDM
 coupling strength between light neutrino $\nu_j$ and heavy Majorana neutrino $N_k$.}
 \label{fig:emlepto_lambda_5D}
 \end{figure}
The Feynman diagrams which will quantify the EMDM coupling strength
are shown in Fig.\ref{fig:emlepto_lambda_5D}.  In Ref.\cite{bell},
no concrete model was suggested wherein the numbers required for
successful leptogenesis could arise naturally. The main motivation
of this paper is to show that it is possible to construct a simple
extension of the SM, where it will be possible to calculate this
effective coupling, which will lead to resonant electromagnetic
leptogenesis. It should, however, be noted that, without the resonance
condition, it is not possible to have the correct amount of
leptogenesis in these models in view of the smallness of the
effective couplings.

The effective dimension-5 coupling constant matrix
$\lambda$ can, thus, be
expressed in a simple form under the assumption of almost equal mass
for the particles in the loop ($M_E\sim M_H \sim M_\Sigma \sim
M_{eq})$ as
\begin{equation}
\lambda=- \; \frac{y_{\Sigma}^*\, y_{H}\,\mu_s\,v_D}{64 \,\pi^2\, M_{eq}^3} \ .
\end{equation}
For a representative set of parameters,
namely $M_N \sim $ few TeV, $M_{eq} \sim $ TeV, $y_{\Sigma}=y_H \sim
{\mathcal{O}}(1)$, $\mu_s\sim $ 10 GeV and $v_D=0.1$ GeV, we have
\[
\lambda \sim 10^{-12} \, {\rm GeV}^{-1} \ .
\]
Note that such values are typical for each of the 
terms $\lambda_{jk}$, while the exact values would depend on 
the exact flavour structure. However, large hierarchies and/or texture 
zeroes are unexpected.

Now we shall investigate the viability of electromagnetic
leptogenesis. We must first check that the out-of-equilibrium decay of
the RH neutrinos can give rise to a nonzero $CP$ asymmetry under the
most general situations. In addition, it is also necessary to examine
whether the parameters considered in our model can produce an
asymmetry of the correct magnitude via the dimension-five dipole
moment operator through the self-energy enhancement.
\section{Resonant Electromagnetic Leptogenesis}
\label{sec:reslepto}
As has been described above, leptogenesis, in this scenario, is driven
by the electromagnetic dipole moment terms appearing in the effective
Lagrangian. Specifically, the lepton asymmetry is generated by the
CP-violating decays of heavy singlet neutrinos to the SM-like light
neutrinos and a photon. As should be apparent from the discussion in the
last section, the size of the EMDM that is generated and the extent of
CP-violation in them is inadequate for thermal leptogenesis. Indeed,
this is a generic problem for all models of electromagnetic
leptogenesis that seek to be consistent with observed physics and yet
be natural.

Given this, we investigate the possibility of a resonant
enhancement. As is well-known, this mechanism is contingent upon the
existence of at least two neutrino species that are very closely
degenerate, and this is what we shall assume. Aesthetically, the
extent of degeneracy needed may seem uncomfortable.  While it can, in
principle, be motivated on the imposition of additional global
symmetries, it should be noted that, in all models of resonant
leptogenesis, the subsequent breaking of the same would, naturally,
lead to a lifting of the degeneracy by a degree that negates the
conditions for resonant enhancement. Hence, rather than introduce
additional symmetries and a host of fields for additional mechanisms
of compressing the spectrum adequately, we just assume that the said
heavy neutrinos are highly degenerate. We will return 
to this point later in this section.

The key quantity of interest is the CP-asymmetry for the decay of
$N_k$ to a photon and a light neutrino given by
\begin{equation}
 \varepsilon_{k} \equiv \frac{
 \Gamma (N_k \rightarrow \nu\,\gamma) -{\Gamma}(N_k\rightarrow \overline{\nu}\,\gamma)}
 {\Gamma (N_k\rightarrow \nu\,\gamma) + {\Gamma} (N_k \rightarrow \overline{\nu}\,\gamma)}\; .
 \label{eqn:cpasy_defn}
\end{equation}
We begin by calculating the lowest order contribution to the decay
rate $\Gamma (N_k\rightarrow \nu_j\,\gamma)$. Since we are interested
in energy scales above the electroweak phase transition, we shall
identify the light neutrino $\nu$ to be a massless left-handed
(SM-like) state while $N's$ are assumed to have Majorana mass of
around 1 TeV. Driven by the effective Lagrangian of
eqn.(\ref{eqn_emdm_lag_eff}), the lowest order decay rate is, thus,
given by
\begin{eqnarray}
 \Gamma(N_k\rightarrow \nu\,\gamma)
  &=& \frac{(\lambda^\dagger \lambda)_{kk}}{4 \pi} \, M_{k}^3 \ ,
    \label{eqn:emdm_decay_rate}
\end{eqnarray}
where all species of (massless) neutrinos $\nu_i$ have been 
summed over.
For effectively creating a lepton asymmetry of the universe, the decay
of, say $N_1$, should be out of equilibrium, the necessary condition
for which is described by $\Gamma(N_1) \lesssim H(T)
\left|_{T=M_1}\right.$ where $H(T)=1.67\, g^{1/2}_{\ast} \, T^{2} 
    / M_{\textrm{Pl}}$ 
is the Hubble parameter at that particular epoch with the Planck
mass $M_{\textrm{Pl}} \simeq 1.2 \times 10^{19}\,\textrm{GeV}$ and the
number of relativistic degrees of freedom $g_{\ast} \simeq 100$. With
the operative temperature $T \sim M_1$, we then have
\begin{equation}
\frac{\left(\lambda^{\dagger}\,\lambda \right)}{4\pi}\,M_1^{3}
\lesssim 1.67\, g^{1/2}_{\ast} \frac{M_1^{2}}{M_{\textrm{Pl}}} \ .
\end{equation}
This is satisfied by the effective EMDM coupling $\lambda$ with $M_1
\sim$ few TeV, for the choice of parameters we have considered here.

\begin{figure}[htb]
 \begin{center}
  \includegraphics[width=15cm,height=3.5cm]{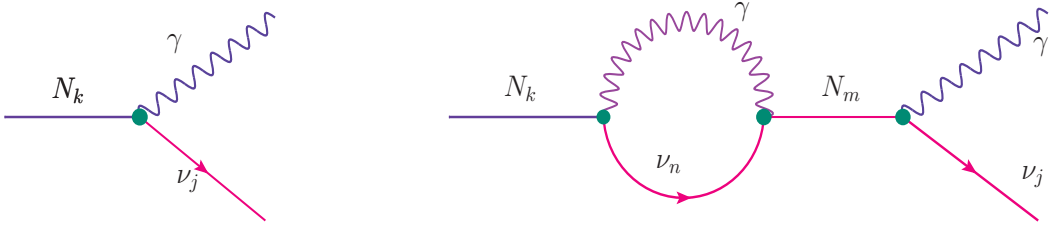}
 \end{center}
 \caption{Self energy diagrams which contribute to the CP-asymmetry of $N_k$ decays via the
 interaction of eqn.(\ref{eqn_emdm_lag_eff}).}
 \label{fig:5D_self}
 \end{figure}

The next task is to calculate the interference
terms between the tree level process and the one-loop diagrams
with on shell intermediate states as shown in Fig.\ref{fig:5D_self}. 
The usual contributions to lepton asymmetry
coming from vertex diagram is found to be very small, i.e,
$\varepsilon_1 =(\lambda^2/4 \pi) M^3_1 \sim (10^{-23} \, {\rm GeV}^{-2})\,
  M^3_1 \sim 10^{-14}$ when $M_1$ is at the TeV scale
and, hence, can be neglected. So, the self
energy contribution will only be considered during the rest of the
discussion.  

The CP-asymmetry here is of a slightly different nature 
as compared to that in standard Yukawa mediated resonant leptogenesis 
\cite{res97,res05,res03,liu}.
 The CP-asymmetry \cite{bell} of $N_k$ decays via the interaction of
eqn.(\ref{eqn_emdm_lag_eff}) has been calculated for the case of
a hierarchical RH neutrino.  In this work, we have calculated the
self-energy diagrams for nearly degenerate heavy RH neutrinos and, in
this case, the CP-asymmetry is found to be
\begin{equation}
\begin{array}{rcl}
 \varepsilon_k & = & \displaystyle 
  \frac{- M^3_k}{2 \pi \, (\lambda^\dagger \lambda)_{kk}} \;
   \sum_{m \neq k} {\rm Im}\left[(\lambda^\dagger \lambda)^2_{k m} \right] \;
 \frac{(M^2_k-M^2_m) M_m}
                  {(M^2_m-M^2_k)^2+M^2_k\,\Gamma^2_m}
\\[3ex]
               &= & \displaystyle 
                 \frac{-2 \, M_k^3}{(\lambda^\dagger \lambda)_{kk}} \;
\sum_{m \neq k} \frac{{\rm Im}\left[(\lambda^\dagger \lambda)^2_{k m} \right]}
                  {M_m^2 \, (\lambda^\dagger \lambda)_{m m}} \;
                  \frac{(M^2_m-M^2_k) \Gamma_m}{(M^2_m-M^2_k)^2+M^2_k\,\Gamma^2_m} \ ,
\end{array}
\label{eqn:eps_Mk} 
\end{equation}
where the expression for the total width $\Gamma_m$ has been 
used to get to the second line. Consider
the case where $M_1\sim M_2 \ll M_3$. From eqn.(\ref{eqn:emdm_decay_rate}), 
it is clear that $\Gamma_1 \sim \Gamma_2$
for nearly degenerate right handed neutrinos with masses $M_{1,2}$.
Hence, $\Gamma_2 \approx
\Gamma_1=(\lambda^\dagger \lambda)_{22} \, M_{2}^3 / / (4 \, \pi)$ 
and, for the $N_1$--dominated case, the CP asymmetry is
\begin{align}
\varepsilon_1 &
              = -\frac{M^2_1}{2 \pi} \; 
   \frac{\sum_{m \neq 1} {\rm Im}\left[(\lambda^\dagger \lambda)^2_{1m} \right]}
               {(\lambda^\dagger \lambda)^2_{11}} \;
              \frac{(M^2_2-M^2_1) M_1 M_2}{(M^2_2-M^2_1)^2+M^2_1\,\Gamma^2_2} \ .
\label{eqn:eps_M1}
\end{align}
As $\Gamma_2 \ll |M_1-M_2| $, even when the heavy neutrinos are quite
degenerate, this further simplifies to
\begin{align}
\varepsilon_1 &
              \approx -\frac{M^2_1}{2 \pi} \;
 \frac{\sum_{m \neq 1} {\rm Im}\left[(\lambda^\dagger \lambda)^2_{1m} \right]}
               {(\lambda^\dagger \lambda)^2_{11}} \;
 \frac{M_1 M_2}{M^2_2-M^2_1} \ .
\label{eqn:eps_M1_simpler}
\end{align}
Clearly, in the almost degenerate case, $\varepsilon_1$ is resonantly 
enhanced. Indeed, writing $M^2_2-M^2_1 \sim 2 M_2(M_2-M_1)$, we have
\begin{align}
\varepsilon_1 &
               \approx \frac{- M^2_1}{4 \pi} \; 
       \frac{\sum_{m \neq 1} {\rm Im}\left[(\lambda^\dagger \lambda)^2_{1m} \right]}
               {(\lambda^\dagger \lambda)^2_{11}} \;
{\cal R}
\label{eqn:eps_M1_final}
\end{align}
where ${\cal R} \equiv M_1 / |M_1-M_2|$. 

As described above, a non-zero $\varepsilon_1$ can give rise to a net
lepton number asymmetry in the Universe, provided its expansion rate
is larger than the decay rate of $N_1$.  The nonperturbative sphaleron
interaction may partially convert this lepton number asymmetry into a
net baryon number asymmetry \cite{bary},
\begin{align}
 \eta_B &
         \simeq -2.96 \times 10^{-2}\,\varepsilon_1\, k
         \nonumber
\end{align}
where $k$ is the efficiency factor measuring the washout effects
associated with the out-of-equilibrium decays of $N_1$.  In our model,
$k \sim {\mathcal{O}}(10^{-3})$.  We, thus, need $|\varepsilon_1| \sim
10^{-5}$ to generate the requisite baryon asymmetry in the Universe.
This is achieved if $|M_2-M_1| \lapp 10^{-7}$ GeV where the mass of
the right handed Majorana neutrinos is around TeV scale.

While such a small mass difference may seem unnatural, it need not be
so. To start with, let us assume that some symmetry forces them to be
exactly degenarate at the tree level. The question of interest, then,
is the extent to which this degeneracy is lifted by quantum
corrections.  To this end, consider a diagram with a vertex
$\lambda_{H D} (D^\dagger D) (H^\dagger H)$ attached to the singly
charged scalar $H$ which runs in the loop contributing to the neutrino
mass.  This engenders a finite contribution to the mass and the
consequent splitting is
\begin{equation}
{\Delta M}_R  
               \sim \frac{\lambda_{H D} \, y^*_H \, y_H}{(4 \pi)^2} \;
                \frac{\langle D \rangle^2}{4 M_E}
\end{equation}
Since $M_E\sim {\cal O}(1 \, \hbox{TeV})$, $v_D = \langle D \rangle
\sim {\cal O}(0.1 \hbox{GeV})$ and $y_H \sim {\cal O}(1)$, a moderate
value of $\lambda_{H D}$ will generate the requisite mass splitting.

Before closing, it may be instructive to make a comparison with the
standard leptogenesis scenario where the CP asymmetry is generated via
the decay $N_R\rightarrow \ell \varphi$, where $\varphi$ denotes a
generic scalar. The decay rate is given by
\[
\Gamma_{\rm standard} = \frac{y^2 \, M_N}{4 \, \pi} \,
                       \left(1 - \, \frac{m_\varphi^2}{m_N^2} \right)^2
\]
with $y$ being the relevant Yukawa coupling. To have leptogenesis
proceed dominantly via the electromagnetic decay, the above decay
rate should be smaller than the corresponding rate into the
electromagnetic channel. This requires $y\lapp 10^{-8}$ for $M_1
\sim {\mathcal{O}}$(few TeV).  In the present case, $y$ would refer
to the effective $\bar N \, \ell \, \Phi$ coupling. Since this is
generated only through $V_{soft}$, we have (for $\mu_{soft} \, \sim
\, 10$ GeV and $m_D \, \sim \, 10$ TeV)
   \[
    y_{\rm eff} \approx h_D \, (\mu^2_{soft} / m_D^2) \sim 10^{-9},
    \]
which is consistent with the value of $y$ estimated above.  However,
this does introduce some amount of fine tuning in the model and the
parameters have to be marginally adjusted to allow the electromagnetic
leptogenesis, a fact that we believe is generic to any realistic model
of electromagnetic leptogenesis.

\section{Summary}\label{sec:summ}
The idea of electromagnetic leptogenesis is a very interesting and
appealing alternative to the standard scenario of leptogenesis. We
have shown that it is indeed possible to have a viable model for
leptogenesis proceeding through such a channel.  However, there are
several generic problems associated with the construction of any model
for electromagnetic leptogenesis. Highlighting these problems, we
showed that the choice of parameters has to be a fine tuned one in the
sense that deviations from the values chosen may not lead to
successful predictions. Also, to have leptogenesis proceed via the
electromagnetic decay channel rather than the standard channel
involving the $N \ell \phi$ Yukawa coupling, it is necessary to have
the Yukawa coupling highly suppressed. This, in a way, leads to some
additional fine tuning the stability of which under radiative
corrections is slightly suspect.  Moreover, the model works only if
there is resonant enhancement of the asymmetry, which requires almost
degenerate heavy neutrinos.  While this might seem an additional fine
tuning, it is not quite so, as it is essentially the same as that
responsible for the heavy neutrino decaying electromagnetically rather
than to a scalar.  Nonetheless, such a fine tuning, perhaps, is a
generic feature of any realistic model of electromagnetic leptogenesis
and warrants a study in its own right.  In spite of all the fine
tuning of parameters, it is difficult to have any reasonable choice of
parameters that 
allows for immediate detection of new physics at the LHC.
While detection of the heavy lepton ($E$) and some of the scalars
is, in principle, possible once the LHC starts operating at its
design energy and luminosity, it would, nonetheless, need a few 
years of accumulation. The CLIC, on the other hand, would 
stand a very good chance of directly observing these states. 
Deciphering the structure of the theory, unfortunately, 
is likely to prove nearly impossible.



\begin{thebibliography}{15}
\bibitem{mink} P. Minkowski, Phys. Lett. B 67 (1977) 421;\\
M. Gell-Mann, P. Ramond and R. Slansky, in Supergravity, eds. D.Z. Freedman and
P. van Nieuwenhuizen (North-Holland, Amsterdam, 1979);\\
T. Yanagida, in Proc. of the Workshop on the Unified Theory and the Baryon Number
in the Universe, Tsukuba, Japan, 1979, eds. O. Sawada and A. Sugamoto;\\
R. N. Mohapatra and G. Senjanovic, Phys. Rev. Lett. 44 (1980) 912.


\bibitem{yana} M. Fukugita and T. Yanagida, Phys. Lett. B 174 (1986) 45.

\bibitem{dukley} J. Dunkley et al. [WMAP Collaboration], arXiv:0803.0586 [astro-ph].
\bibitem{buch} For recent reviews, see:
 W. Buchmuller, R. D. Peccei and T. Yanagida, Ann. Rev. Nucl. Part. Sci. 55 (2005).
 311 [arXiv:hep-ph/0502169];\\ S. Davidson, E. Nardi and Y. Nir, arXiv:0802.2962 [hep-ph].


\bibitem{kuz} V. A. Kuzmin, V. A. Rubakov and M. E. Shaposhnikov, Phys. Lett. B 155 (1985) 36.


\bibitem{bary} W. Buchm\"uller, P. Di Bari, and M. Plumacher, New J. Phys. 6, 105 (2004);\\
G.F. Giudice, A. Notari, M. Raidal, A. Riotto, and A. Strumia, Nucl. Phys. B 685, 89 (2004).

\bibitem{res97} M. Flanz, E.A. Paschos, U. Sarkar and J. Weiss,
Phys. Lett. B389 (1996) 693;\\
A. Pilaftsis, Phys. Rev. D56 (1997) 5431; Nucl. Phys. B504 (1997) 61.
\bibitem{filf} A. Pilaftsis, Int. J. Mod. Phys. A 14 (1999) 1811
\bibitem{res02} T. Hambye, Nucl. Phys. B633 (2002) 171.
\bibitem{res03} A. Pilaftsis and T.E.J. Underwood, arXiv:hep-ph/0309342.
\bibitem{res04} T. Hambye, J. March-Russell and S.M. West, 
JHEP 0407:070,2004 [hep-ph/0403183].
\bibitem{res05} A. Pilaftsis, Phys. Rev. Lett. 95 (2005) 081602.




\bibitem{liu} J. Liu and G. Segre, Phys. Rev. D 48 (1993) 4609;\\
M. Flanz, E.A. Paschos and U. Sarkar, Phys. Lett. B 345 (1995) 248;\\
L. Covi, E. Roulet and F. Vissani, Phys. Lett. B 384 (1996) 169.

\bibitem{bell} N. Bell, B. Kayser, and S. Law, Phys. Rev. D 78, 085024 (2008).

\bibitem{vol} M.B. Voloshin, Yad. Phys. 48 (1988) 804 [Sov. J. Nucl. Phys. 48].


\bibitem{kim} J.E. Kim, Phys. Rev. {\bf D~14} (1976) 3000.\\
B.W. Lee and R.E. Shrock, Phys. Rev. {\bf D~16} (1977) 1444.\\
W.J. Marciano and A.I. Sanda, Phys. Lett. {\bf B~67} (1977) 303.

\bibitem{valle} J. Schechter, J.W.F. Valle, Phys. Rev. D {\bf 24}, 1883 (1981);
  Phys. Rev. D {\bf 25}, 283 (1982);\\ J.F. Nieves, Phys. Rev. D {\bf 26}, 3152 (1982);\\
  B. Kayser, Phys. Rev. D {\bf 26}, 1662 (1982);\\ R.E. Schrock, Nuc. Phys. B {\bf 206},
  359 (1982);\\ L.F. Li and F. Wilczek, Phys. Rev. D {\bf 25}, 143 (1982);\\
D.~Choudhury and U.~Sarkar, Phys. Lett. {\bf B235}, 113 (1990).



%
\end{thebibliography}
\end{document}